\documentclass[amssymb,aps,floatfix,floats,preprintnumbers,twocolumn,prb,showpacs]{revtex4}
\usepackage{rcs}
\usepackage{color,graphicx}
\usepackage{dcolumn}
\usepackage{bm}
\usepackage[normalem]{ulem}
\usepackage{chngcntr}
\usepackage{natbib}
\usepackage[thinspace,thinqspace]{SIunits}
\usepackage{lipsum}
\usepackage{float}
\usepackage{amsmath}
\usepackage{lmodern}
\usepackage{physics}
\usepackage[a4paper, total={180mm,260mm} ]{geometry}

\begin{document}

\title{Large positive correlation between the effective electron mass and the multipolar fluctuation in the heavy-fermion metal Ce$_{1-x}$La$_x$B$_6$}

\author {D. J. Jang$^{1,}$\footnote[2]{Corresponding author\\ e-mail: dongjin.jang@cpfs.mpg.de}, P. Y. Portnichenko$^2$, A. S. Cameron$^2$, G. Friemel$^3$, A. V. Dukhnenko$^4$, N. Y. Shitsevalova$^4$, V. B. Filipov$^4$, A. Schneidewind$^5$, A. Ivanov$^6$, D. S. Inosov$^2$ and M. Brando$^{1}$
}

\affiliation{$^1$Max Planck Institut f{\"u}r Chemische Physik fester Stoffe, 01187 Dresden, Germany}
\affiliation{$^2$Institut f{\"u}r Festk{\"o}rper- und Materialphysik, TU Dresden, 01069 Dresden, Germany}
\affiliation{$^3$Max Planck Institut f{\"u}r Festk{\"o}perforschung, 70569 Stuttgart, Germany}
\affiliation{$^4$I. M. Frantsevich Institute for Problems of Materials Science of
NAS, 3 Krzhyzhanovsky Street, Kiev 03680, Ukraine}
\affiliation{$^5$J{\"u}lich Center for Neutron Science (JCNS), Forschungszentrum J{\"u}lich GmbH,\\ \mbox{Outstation at Heinz Maier\,--\,Leibnitz Zentrum (MLZ), Lichtenbergstra{\ss}e 1, D-85747 Garching, Germany}}
\affiliation{$^6$Institut Laue-Langevin, 71 avenue des Martyrs, CS 20156, 38042 Grenoble cedex 9, France}

\date{\today}

\begin{abstract}
For the last few decades, researchers have been intrigued by multipolar ordering phenomena while looking for the related quantum criticality in the heavy-fermion Kondo system Ce$_{1-x}$La$_{x}$B$_6$. However, critical phenomena induced by substitution level ($x$), temperature ($T$), and magnetic field ($B$) are poorly understood despite a large collection of experimental results is available. In this work, we present $T$-$B$, $x$-$T$, and $x$-$B$ phase diagrams of Ce$_{1-x}$La$_x$B$_6$ ($\mathbf{B}\parallel[110]$). These are completed by analyzing heat capacity, magnetocaloric effect (MCE), and elastic neutron scattering. A drastic increase of the Sommerfeld coefficient $\gamma_0$, which is estimated from the heat capacity down to 0.05~\,K, is observed with increasing $x$. The precise $T$-$B$ phase diagram which includes an unforeseen high-entropy region is drawn by analyzing the MCE for the first time in Ce$_{1-x}$La$_x$B$_6$. The $x$-$B$ phase diagram, which supports the existence of a QCP at $x>0.75$, is obtained by the same analysis. A detailed interpretation of phase diagrams strongly indicates positive correlation between the fluctuating multipoles and the effective electron mass.
\end{abstract}

\maketitle 
\section{Intronduction}
The sixfold degenerate wave functions of an isolated Ce$^{3+}$ ion are split into the $\Gamma_8$ quartet and the $\Gamma_7$ doublet under the crystal electric field with octahedral symmetry in Ce$_{1-x}$La$_x$B$_6$. The energy difference between the ground-state (GS) quartet and the excited doublet is about 46~\,meV (Ref.~\,1). Hence, low-temperature ($T$) and low-field ($B$) properties are governed by the $\Gamma_8$ quartet and there are possibilities to observe multipolar moments including a dipole, a quadrupole, and an octupole (atomic 4$f$ current distribution with zero magnetic moment). What has made this material so intriguing for the last few decades is that the above mentioned multipoles can interact through superexchange~\cite{Ohkawa1,Ohkawa2,Shiina}. Indeed, antiferromagnetic (AFM)~\cite{Effantin,Hiroi,Zaharko}, antiferroquadrupolar (AFQ)~\cite{Sera,Sera1,Nakao,Takigawa,Sakai}, and antiferrooctupolar (AFO) orderings~\cite{Sera,Matsumura,Matsumura1,Mannix,Kuwahara} have been observed, and phase transitions between these ordered phases have also been intensively studied. By convention, the spin/pseudospin (orbital) paramagnetic (PM) phase, the AFQ phase, and the AFM phase are referred to as phases I, II, and III, respectively. The phase IV which appears around $x\simeq0.3$ is suspected to represent AFO order~\cite{Kuwahara}. 

However, even though largely scattered information about the multipolar phase transitions is gathered~\cite{Alistair}, the result is rather difficult to understand. The $x$-$T$ phase diagram  at $B=0$ reveals a weakening of the phase IV as $T$ is increased, while the $x$-$B$ phase diagram at $T$=0 reveals an enhancement of the same phase as $B$ is increased. Therefore, a discontinuity of the phase IV has been introduced along the $x$-axis in the up to date $x$-$T$-$B$ phase diagram of Ce$_{1-x}$La$_x$B$_6$, and the criticality at the verge of the phase IV is poorly understood.

In this article, we investigate physical properties of Ce$_{1-x}$La$_x$B$_6$ ($x$=0, 0.18, 0.23, 0.28, 0.5, 0.75) by measuring heat capacity at ambient pressure, magnetic Bragg intensity, and the magnetocaloric effect (MCE) with $\mathbf{B} \parallel [110]$. $C_p$ and elastic neutron scattering (NS) measurements are analyzed to construct $x$-dependent $T$-$B$ phase diagrams and the $x$-$T$ phase diagram which show intricate transitions between multipolar phases. From the low $T$ analysis of the $C_p/T$, the Sommerfeld coefficient, $\gamma_0$, is extracted. The $B$-dependent $\gamma_0$ exhibits local maxima at critical points. At low $B$, the overall magnitude of $\gamma_0$ drastically increases for $x>0.28$. Based on the unprecedented analysis of the MCE in this system, a novel high-entropy region in the $x$-$B$ plane is revealed. This region expands as $x$ is increased until it merges with the phase IV. Most importantly, we present an understandable $x$-$B$ phase diagram, which shows weakening of the phase IV as $x$ is increased and predicts the emergence of a putative quantum critical point (QCP) at the verge of the phase IV. The new observations concerning the variation of $\gamma_0$ on the $x$-$B$ plane are explicable by a large positive correlation between multipolar fluctuations and the effective electron mass, $m^*$, within heavy Fermi-liquid (FL) description.
%
\begin{figure*}[ht]
\includegraphics[width=\textwidth]{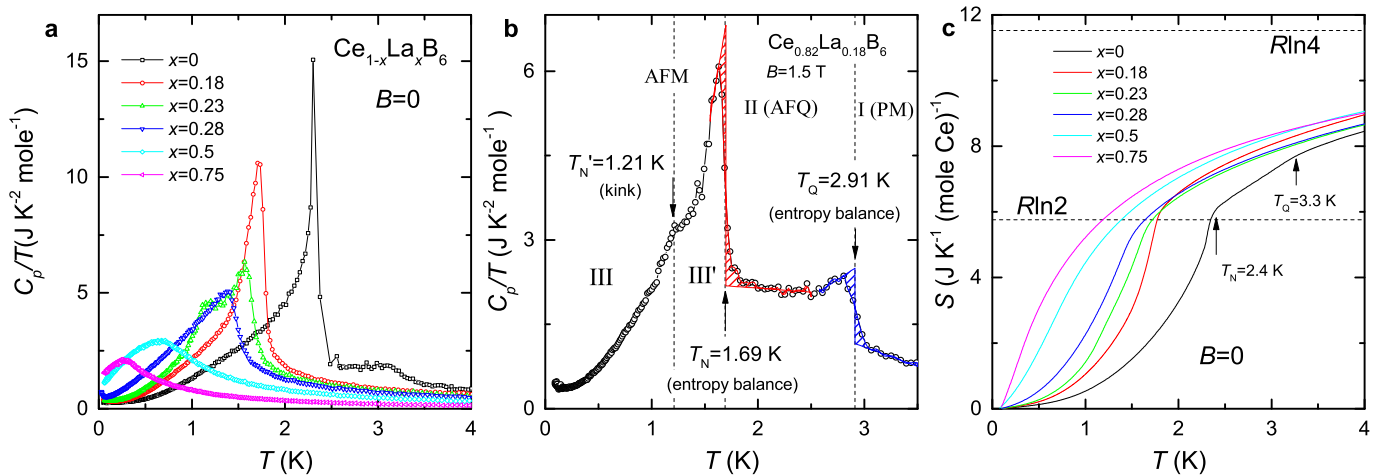}
\caption{(color) {\bf Specific heat capacity $C_p(T)$ and entropy $S_p(T)$ with  {\bf \textit{x}~\,=~\,0}, 0.18, 0.23 0.28, 0.5, 0.75.} ({\bf a}) $T$-dependent specific heat functions with different values of $x$ and for 0.05~\,K $\leq T \leq4$~\,K. ({\bf b}) Representative $C_p(T)/T$ curve obtained from Ce$_{0.82}$La$_{0.18}$B$_6$ for applied field of $B=1.5$~\,T. Phase I is a spin/pseudospin PM phase, phase II is AFQ phase, phase III is a non-colinear AFM phase, and phase III$^\prime$ is a collinear AFM phase. To find the MF transition temperature, we used method of entropy balance resulting in $T_\textrm{N} =1.69$~\,K (red line and red hatched area) and $T_\textrm{Q}=2.91$~\,K (blue line and blue hatched area). The transition between AFM phases is indicated by the kink at $T_\textrm{N}^\prime=1.21$~\,K. ({\bf c}) Entropy curves per 1 mole of Ce atom are exhibited for samples of different values of $x$.}
\label{fig1}
\end{figure*}
%
\section{Results}
In Fig.~\ref{fig1}a, phase transitions in Ce$_{1-x}$La$_{x}$B$_6$ are signaled by sharp anomalies in $C_p/T$ if the Ce concentration is high. However, the anomalies are weakened and the overall magnitude of $C_{p}(T)/T$ is diminished with increasing La concentration. To define the critical temperature, it should be reminded that the mean-field (MF) theory predicts a discontinuous jump in $C_p/T$ at the transition temperature. We can extrapolate the discontinuous jump as noted by the red and blue lines drawn in Fig.~\ref{fig1}b. Here, the so-called method of entropy balance is applied to define the N\'{e}el temperature $T_\textrm{N}$ and the AFQ transition temperature $T_\textrm{Q}$: Note that hatched regions with the same color have the same area. On the other hand, a transition from a non-collinear AFM phase (III) to a collinear AFM phase (III$^\prime$) at $T_\textrm{N}^\prime$ typically results in a kink in $C_p/T$. Each region of multipolar phases and the corresponding transitions are confirmed from the literature~\cite{Alistair,Friemel}. Fig~\ref{fig1}c shows that the entropy as a function of $T$ loses its features for the phase transitions (see kinks in the black curve at $T_\textrm{N}=2.4$~\,K and $T_\textrm{Q}=$3.3~\,K) as $x$ is increased. Meanwhile, the overall magnitude of the entropy is increased with $x$. Due to the remaining short-range interactions between multipoles in phase I, we can observe saturated entropy of the randomly populated quartet only well above 10~\,K (Ref. 19). The field is applied along [110] crystallographic orientation because a brief $x$-$T$-$B$ phase diagram is only available for this direction as a reference~\cite{Alistair,Friemel}. 

%
\begin{figure*}[ht]
\includegraphics[width=\textwidth]{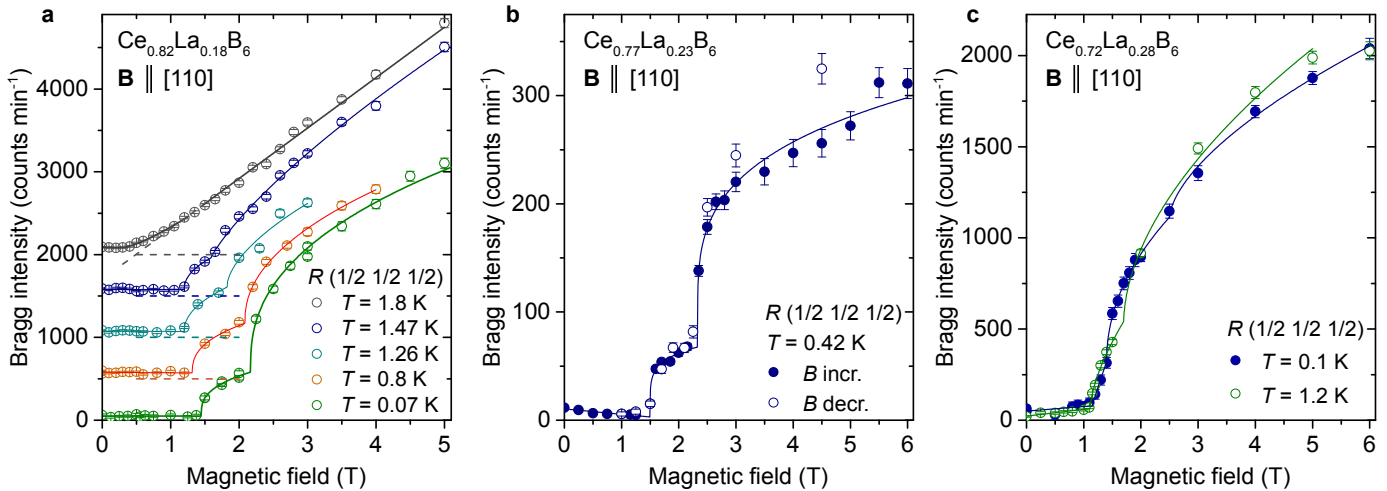}
\caption{(color) {\bf $B$- and $T$-dependent (1/2 1/2 1/2) Bragg intensity from Ce$_{1-x}$La$_{x}$B$_6$ with {\bf \textit{x}}~\,=~\,0.18, 0.23, and 0.28.} ({\bf a}) Field-induced (1/2 1/2 1/2) Bragg intensity measured from Ce$_{0.82}$La$_{0.18}$B$_6$ with different temperatures are shown. Each curve is shifted along the $y$ axis for increased visibility. The lowest lying curve is for the lowest $T$. ({\bf b}) Field-induced (1/2 1/2 1/2) Bragg intensity from Ce$_{0.77}$La$_{0.23}$B$_6$ with $T=0.42$~\,K. Filled-circles designate neutron counts in increasing $B$. Hollow-circles designate neutron counts in decreased $B$. ({\bf c}) Field-induced (1/2 1/2 1/2) Bragg intensity obtained from Ce$_{0.72}$La$_{0.28}$B$_6$ at two different temperatures. Filled-circles and hollow circles denote the neutron counts at $T=0.1$~\,K and $T=1.2$~\,K, respectively. $\mathbf{B}\parallel [110]$ in all of the measurements.}
\label{fig2}
\end{figure*}
%

Figure~\ref{fig2} shows the field-induced Bragg intensity at the (1/2 1/2 1/2) wave vector in three samples of Ce$_{1-x}$La$_x$B$_6$ measured by NS. Unlike the AFM phase~\cite{Effantin,Hiroi,Zaharko,Tayama}, which exhibits Bragg scattering in zero magnetic field, the hidden-order phase II can only be visualized by the elastic intensity that originates from the field-induced dipolar moments modulated by the underlying AFQ structure. Not only does this offer a method to reveal the presence of AFQ correlations by NS, but it may also provide information about the type of the multipoles involved~\cite{Pavlo}.

The field dependence in Fig. 2 reveals that the onset of intensity measured at (1/2 1/2 1/2) in all the studied samples occurs in several steps. In Ce$_{0.82}$La$_{0.18}$B$_6$ (Fig.~\ref{fig2}a), the intensity remains almost zero within the phase III, but starts to increase already at $B=1.5$~T inside the phase III$^\prime$ for $T=0.07$~\,K. This demonstrates that this distinct phase represents a combination of conventional (dipolar) AFM and AFQ order parameters, in contrast to the purely dipolar phase III. Upon entering the phase II, an even steeper jump in intensity is observed around 2.2~\,T. With increasing temperature, both transitions are shifted towards lower fields as the AFM phases are suppressed. A similar two-step increase without field-history dependence is also seen in the low-temperature measurement on Ce$_{0.77}$La$_{0.23}$B$_6$ (Fig.~\ref{fig2}b). In our third sample, Ce$_{0.72}$La$_{0.28}$B$_6$, the zero-field GS is reportedly taken over by the phase IV (Ref.~\,18), but the phases III and III$^\prime$ can still be stabilized by the application of field. As a result, the corresponding field dependence in Fig.~\ref{fig2}c now exhibits a series of three steps at $T=0.1$~K: a broadened transition near 1~T from phase IV to III, a kink around 1.5~T corresponding to the III-III$^\prime$ transition, and the third transition to phase II at 2.5~\,T (Ref.~\,18). We note that for the La substitution level of $x=0.28$, some initial increase in field-induced intensity is observed already within phase III, unlike at low La concentrations where no signal within phase III was found. This suggests that the field-stabilized phase III in La-substituted samples is different from the zero-field phase III with regard to its intermixing with the AFQ order parameter. On the other hand, within phase IV, i. e. for fields below 1~\,T, the (1/2 1/2 1/2) intensity is nearly insensitive to the field.

\begin{figure*}[ht]
\includegraphics[width=\textwidth]{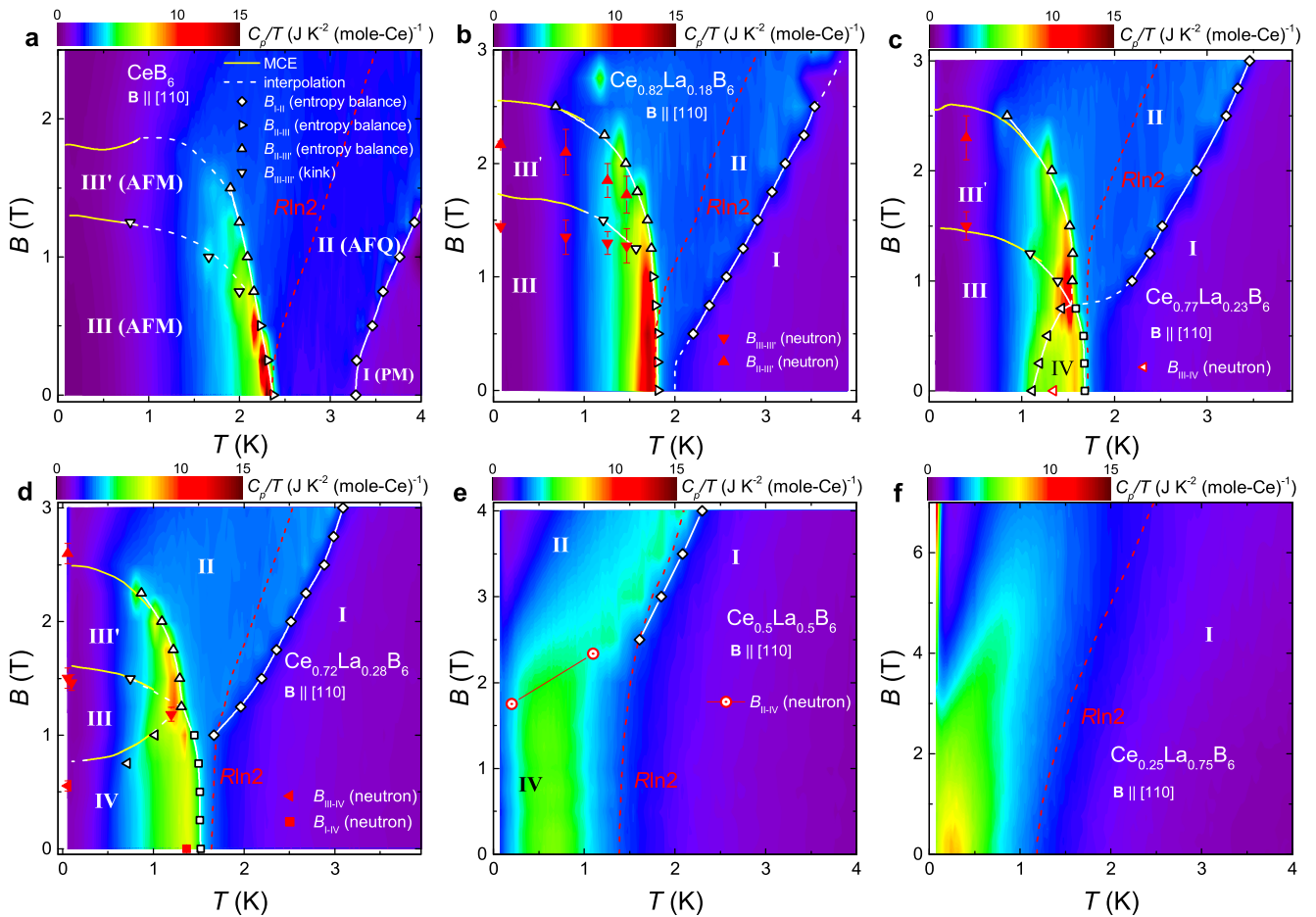}
\caption{(color) {\bf $\textit{T}$-$\textit{B}$ phase diagrams of Ce$_{1-x}$La$_x$B$_6$ with applied fields parallel to ${\bf \left[110\right]}$ direction.} ({\bf a}) The color plot is interpolated from $C_{p,B}(T)/T$ of CeB$_6$. Transition temperatures are determined by an entropy balance or by a notable kink in $C_{p,B}(T)/T$. In the $T$-$B$ plane, critical fields are estimated as a function of $T$ (see the legend). ({\bf b}) The color plot is interpolated from $C_{p,B}(T)/T$ of Ce$_{0.82}$La$_{0.18}$B$_6$. The critical fields determined from the changes in the (1/2 1/2 1/2) Bragg intensity are superimposed. Upward triangles in red denotes the II-III transition and downwards triangles in red denotes the III-III$^\prime$ transition. ({\bf c}) When $x=0.23$, phase II exists only if a field is applied and phase IV appears over a narrow low $T$ region. A change of Bragg intensity is also observed when the system enters to phase III from phase IV as marked by the left-triangle with red-edge. ({\bf d}) For $x=0.28$, phase IV becomes the GS and the red square notes that there is a change in magnetic Bragg intensity. The left-triangle in red denotes $B_\textrm{III-IV}$ taken from Fig.~\ref{fig2}c. ({\bf e}) When $x=0.5$, phase III is absent and a broad phase transition between phases I and IV is remained at low $B$. (1/2 1/2 1/2) Bragg intensity is induced by field upon entering phase II from phase IV (see dotted-circles with red edge). ({\bf f}) As $x$ is increased to 0.75, we cannot find a notable sign for a distinct phase transition except rather sharp features at the lower-left corner. Red dashed-lines in all of the panels represent isentropic lines with $S_p=R\textrm{ln}2$.}
\label{fig3}
\end{figure*}
\begin{figure*}[ht]
\includegraphics[width=\textwidth]{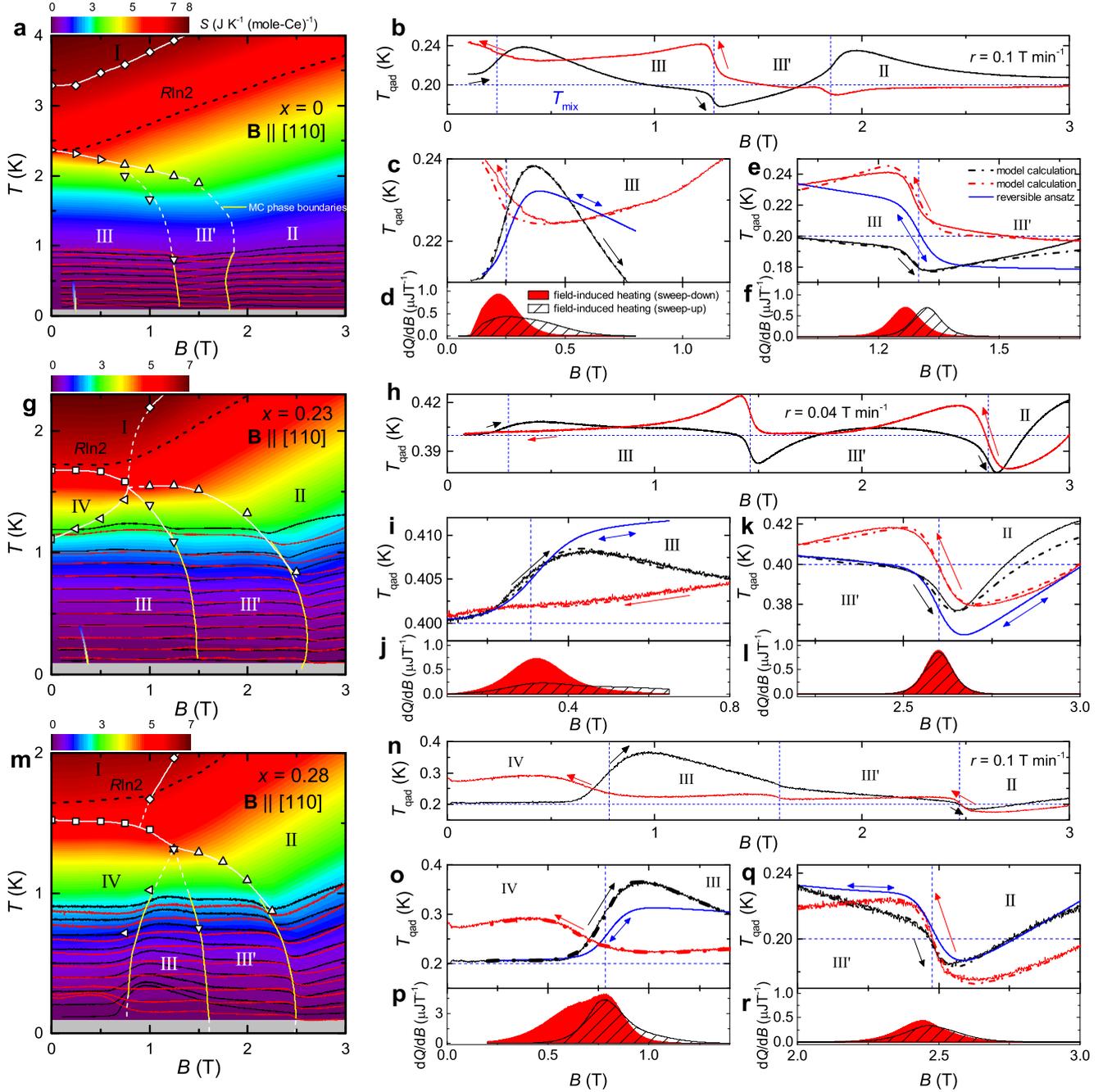}
\caption{(color) {\bf {\bf $\textit{S}_p\left(\textit{B},\textit{T}\right)$}, {\bf $\textit{B}$-$\textit{T}$} phase diagrams, and quasi-adiabatic MC anomalies in Ce$_{1-x}$La$_x$B$_6$ with applied fields parallel to ${\bf \left[110\right]}$ direction.} ({\bf a}) The background is the contour plot of $S_p(B,T)$ from CeB$_6$. Yellow solid lines denote phase boundaries determined by MC anomalies. Black (red) line is $T_\textrm{qad}(B)$ in up-(down-) sweep with sweep-rate $r$=0.1~\,T min$^{-1}$. The short line with color gradient at low field and temperature emphasizes unclosed borderline under which the entropy is higher than that of phase III. ({\bf b}) One of the $T_\textrm{qad}(B)$ curves is magnified from the panel {\bf a}. $T_\textrm{mix}$ is fixed at 0.2~\,K. ({\bf c}) $T_\textrm{qad}(B)$ curves below 1.2~\,T are enlarged from panel {\bf b}. The blue solid line is the reversible ansatz. The black and red dash-dotted lines are calculated by the numerical model explained in the text and the Appendix C. ({\bf d}) The envelopes of the black-hatched and the red-filled areas denote distributions of the field-induced heat $dQ(B)/dB$ in sweep-up and down, respectively. ({\bf e}) Region around the III-III$^\prime$ boundary is magnified from the panel {\bf b}. Results of the model calculations and the reversible ansatz are shown together. ({\bf f}) $dQ(B)/dB$ curves around the III-III$^\prime$ boundary are shown. ({\bf g}) Contour plot of $S_p(B,T)$ and the $B$-$T$ phase diagram of Ce$_{0.77}$La$_{0.23}$B$_6$ are exhibited with $r$=0.04~\,T min$^{-1}$. As in the panel {\bf a}, the high-entropy region is noted by the unclosed borderline with color gradient. ({\bf h}) Curves for $T_\textrm{qad}(B)$ with $T_\textrm{mix}$=0.4~\,K are magnified from the panel {\bf g}. ({\bf i})-({\bf l}) MC anomalies for selected boundaries from the panel {\bf h} are analyzed. ({\bf m}) Contour plot of $S_p(B,T)$ and the $B$-$T$ phase diagram of Ce$_{0.72}$La$_{0.28}$B$_6$ are shown. $r=0.1$~\,T min$^{-1}$. ({\bf n}) $T_\textrm{qad}(B)$ curves with $T_\textrm{mix}=0.2$~\,K are magnified from the panel {\bf m}. ({\bf o})-({\bf r}) MC anomalies around selected phase boundaries from panel {\bf n} are analyzed.}
\label{fig4}
\end{figure*}

$T$-$B$ phase diagrams of Ce$_{1-x}$La$_x$B$_6$ with $\mathbf{B}\parallel[110]$ are displayed in Fig.~\ref{fig3}. Phase boundaries and the degree of critical fluctuations are readily recognizable by the contrasting colors in the contour plots of $C_p(T,B)/T$. The low $T$ phase boundaries (yellow solid lines) are determined from the MCE analysis which will be explained with Fig.~\ref{fig4}. The notations for the critical fields are straightforward, e.g. the critical field across the phases I and II is labelled by $B_\textrm{I-II}$. In Figs.~\ref{fig3}b-\ref{fig3}d, the red upward-triangle and the red downward-triangle denote $B_\textrm{II-III$^\prime$}$ and $B_\textrm{III-III$^\prime$}$ determined by NS shown in Fig.~\ref{fig2}. In phase II, the field-induced AFO order stabilizes the AFQ order resulting in positive slope of $B_\textrm{I-II}(T)$~\cite{Matsumura}. Phase II is gradually weakened and transformed into phase I with increasing $x$.

For $x=0.23$, a certain phase appears in the temperature range of 1~\,K $<T<1.6$~\,K and fields below 1~\,T (Fig.~\ref{fig3}c). When $x=0.28$, this phase becomes the GS (Fig.~\ref{fig3}d). Since the primary order parameter of this phase has not been conclusively determined, it is customarily called phase IV. The nature of phase IV will be discussed later. In Fig.~\ref{fig3}e, broad peaks are almost insensitive to fields below 2~\,T. It will be argued that the phase IV survives in Ce$_{0.5}$La$_{0.5}$B$_6$ with a rather broad heat capacity anomaly. In Fig.~\ref{fig3}f, Ce$_{0.25}$La$_{0.75}$B$_6$ shows field-insensitive peaks only in the low $T$ and low $B$ region. In the rest of the $T$-$B$ plane, a specific-heat anomaly is broadened and moves to the high-$T$ side as $B$ is increased. This is reminiscent of the Kondo impurity behavior, but the detailed shape of $C_p(T,B)/T$ deviates from the resonance-level model~\cite{Schotte,Desgranges}. 

%
%

Now, we present the quasi-adiabatic MCE observed in Ce$_{1-x}$La$_x$B$_6$. The ideal MCE is defined as reversible temperature changes of a thermally isolated specimen controlled by an external magnetic field. Suppose the $S_{p,B_\textrm{i}}(T)$ curve is lying higher than the $S_{p,B_\textrm{f}}(T)$ curve. The sample tends to reduce its entropy by releasing a certain amount of heat as the field is changed from $B_\textrm{i}$ to $B_\textrm{f}$. But the heat must be dissipated and the temperature has to be increased to keep the entropy unchanged. In the reverse procedure, cooling of the sample occurs. Likewise, a phase transition from a magnetically disordered state to an ordered state is described by a sharp increase of the sample temperature and vice versa. In practice, however, the sample is in a quasi-adiabatic condition in which the sample temperature, $T_\textrm{qad}$, slowly and monotonically relaxes to $T_\textrm{mix}$, the temperature of the mixing chamber of the dilution fridge to which the sample is mounted. Moreover, a field-induced heating around a critical region is found to be essential to describe the irreversible $T_\textrm{qad}(B)$ upon field cycling. We have constructed a mathematical model (see Appendix C for more details) for the above observations, and the basic principle is graphically summarized in Fig.~\ref{fig4}. Eddy current heating is negligible in this experiment (Appendix C).

The background of Fig.~\ref{fig4}a represents the color-coded entropy $S_p(T,B)$ of CeB$_6$ calculated from $C_p/T$. Black (red) solid lines indicate $T_\textrm{qad}(B)$ when the field is increased (decreased). In Fig.~\ref{fig4}b, representative magnetocaloric (MC) sweeps with $T_\textrm{mix}=0.2$~\,K and the sweep-rate $r=0.1$~\,T min$^{-1}$ are magnified. To simulate $T_\textrm{qad}(B)$, we assume a continuous phase transition, and start with a reversible ansatz for the true MCE. The distribution of the excessive field-induced heat, $dQ(B)/dB$, is also approximated in terms of an analytic function. Then, these factors are taken into the differential equation for the theoretical quasi-adiabatic temperature and consecutively adjusted until the solution is optimized to $T_\textrm{qad}(B)$. Fig.~\ref{fig4}c is the amplified view of Fig.~\ref{fig4}b at low $B$. Here, the reversible ansatz is the blue solid line. In Fig.~\ref{fig4}d, the envelope of the black hatched-area (red-area) pertains to $dQ(B)/dB$ in sweep-up (sweep-down) mode. The optimized solutions are denoted by dash-dotted lines in Fig.~\ref{fig4}c. Note that the major cause of the irreversibility is indeed the field-induced heating accompanied by the maximum MCE. Magnified view of $T_\textrm{qad}(B)$ near the III-III$^\prime$ boundary is shown in Fig.~\ref{fig4}e. An about two times larger change of $T_\textrm{qad}(B)$ in down-sweep is well reproduced by the hysteretic field-induced heating (Figs.~\ref{fig4}e and ~\ref{fig4}f). The III-III$^\prime$ boundary is also explained by the model (not shown). We tried to simulate the total field-induced heat, $Q$, as closely as possible for both sweep directions, but no constraint regarding $dQ/dB$ was imposed. These conditions reflect hysteretic behaviors across certain phase boundaries observed in magnetic susceptibility, magnetization, and NS~\cite{Hiroi,Friemel,Tayama}. 

The yellow solid lines in Figs.~\ref{fig4}a,~\ref{fig4}g, and ~\ref{fig4}m connect critical points referenced to steepest slopes in the reversible ansatzes (see blue vertical dashed lines in other panels of the Fig.~\ref{fig4}). On the other hand, a line which fades away above $T=0.4$~\,K is inserted in Fig.~\ref{fig4}a and stresses that the MCE is weakened as the temperature is increased. The shape of the reversible ansatz confirms that the magnetic entropy under this line is larger than the magnetic entropy of the well-defined AFM phase. While this unclosed borderline cannot be attributed to a phase transition in the strict sense, AFM domain selection or motion would also be insufficient to explain the origin of this line as these effects usually continue to $T_\textrm{N}$ (Ref.~\,20). 

Fig.~\ref{fig4}g shows the $B$-$T$ phase diagram of Ce$_{0.77}$La$_{0.23}$B$_6$. Curves of $T_\textrm{qad}(B)$ with $T_\textrm{mix}=0.4$~\,K and $r=0.04$~\,T min$^{-1}$ are magnified in Fig.~\ref{fig4}h. The model is applied to explain the weak features around $B=0.4$~\,T (Figs.~\ref{fig4}i and ~\ref{fig4}j). The phase II-III$^\prime$ boundary is also analyzed (Figs.~\ref{fig4}k and ~\ref{fig4}l). The unclosed borderline with the color gradient is shifted to higher fields compared with the one found in CeB$_6$. 

In Fig.~\ref{fig4}m, phase IV becomes the GS. The model is applied to the representative $T_\textrm{qad}(B)$ shown in Fig.~\ref{fig4}n. Analyses of III-IV, III-III$^\prime$ (not shown), and II-III$^\prime$ transitions repeatedly confirm that substantial $Q$ is accompanied by the large MCE (Figs.~\ref{fig4}o~\--~\ref{fig4}r). The III-IV boundary found by the MCE analysis is distinct, but we cannot draw such a well-defined phase boundary by observing a low $T$ AFM structure since it is strongly field-history-dependent~\cite{Friemel}. Hence, the sudden onset of the AFQ phase with field-induced Bragg intensity (see Fig.~\ref{fig2}c) might explain the sharp III-IV boundary.

%
%
\section{Discussion}
Before we discuss our new findings, it has to be emphasized that the MCE analysis has played essential role to present new perspectives on thermodynamic phenomena in Ce$_{1-x}$La$_x$B$_6$. Experimentally, the completion of the precise $B$-$T$ phase diagram is possible because $T_\textrm{qad}(B)$ is extremely sensitive to the entropy change, and this property allows us to explore relevant and the most interesting area in the $B$-$T$ plane where anomalies in $C_p$, NS, ultrasonic attenuation~\cite{Nakamura2}, and thermal expansion~\cite{Akatsu} are not easily detectable. Theoretically, our phenomenological model implies substantial $r$-dependent quantum mechanical friction appears due to strongly fluctuating local magnetic moments, and suggest a non-equilibrium formulation for a field-driven phase transition. In terms of a quasiparticle dynamics, it is suspected that inelastic magnon-magnon and magnon-electron scatterings might be prominent. A detailed study about a $r$-dependent $Q$ and a type of transition is now underway: A first-order phase transition is expected if both $Q$ and $dQ/dB$ are highly irreversible as in the III-IV transition (Fig.~\ref{fig4}p), while a second-order transition is characterized by reversible $Q$ and $dQ/dB$ (Fig.~\ref{fig4}l and ~\ref{fig4}q). A continuous III-III$^\prime$ transition is described by reversible $Q$ and irreversible $dQ/dB$ because of the domain hysteresis (Fig.~\ref{fig4}f). 

%
\begin{figure*}[ht]
\includegraphics[width=\textwidth]{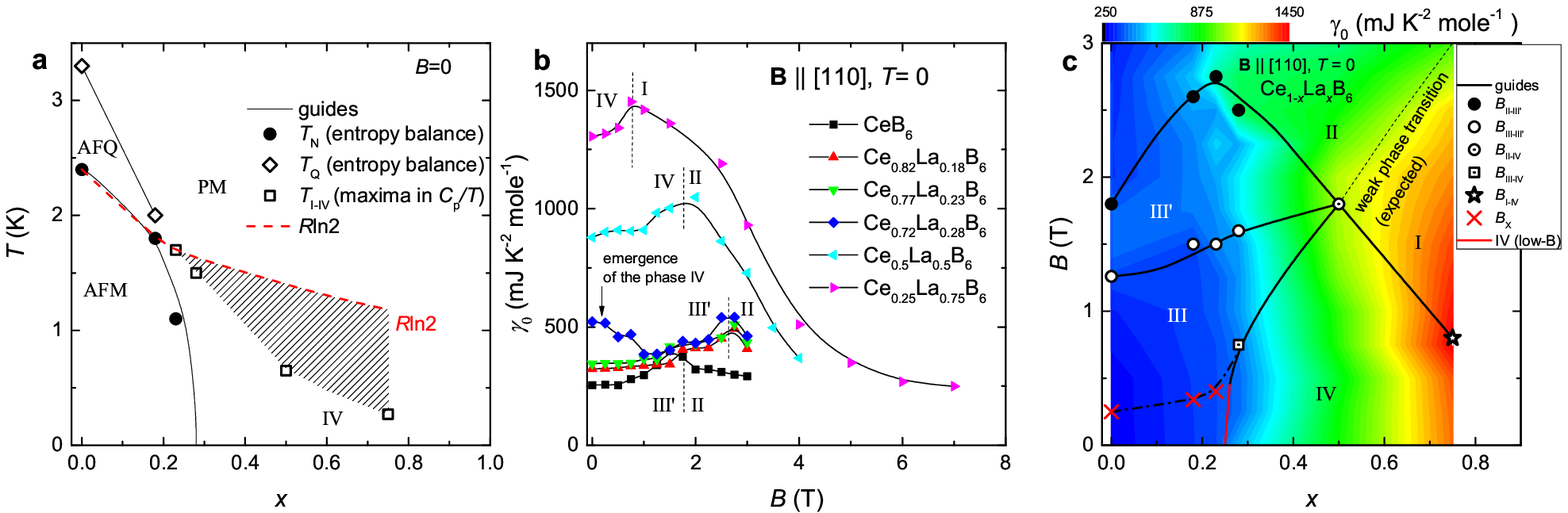}
\caption{(color) {\bf The ${\bf \textit{B}=0}$ {\bf $\textit{x}$-$\textit{T}$} phase diagram, {\bf $\textit{B}$}- and {\bf $\textit{x}$}-dependent {\bf $\gamma_0$}, and ${\bf \textit{T}=0}$ {\bf $\textit{x}$-$\textit{B}$} phase diagram of Ce$_{1-x}$La$_x$B$_6$.} ({\bf a}) $B=0$ $x$-$T$ phase diagram is shown. It is completed by placing coordinates of specific heat anomalies. The I-IV transition with transition temperature $T_\textrm{I-IV}$ becomes broad above $x\simeq0.3$. The region of broad transitions is roughly coincident with the hatched area between the $S_p=R\textrm{ln}2$ isentropic line and the $T_\textrm{I-IV}(x)$. ({\bf b}) The Sommerfeld coefficient as a function of $B$ with different values of $x$ is presented. Locations of local maxima (see short vertical dashed lines) are coincident with the critical points extrapolated from the lowest $T$ MCE anomalies. ({\bf c}) The black dot and the white circle designate $B_\textrm{II-III$^\prime$}$ and $B_\textrm{III-III$^\prime$}$ determined by the MCE analysis, respectively. The high-entropy region exists below $B_\times$ and the black dash-dotted line connects red $\times$ symbols as a guide to the eye. The red segment for the low $B$ III-IV phase boundary is to emphasizes large increase of $\gamma_0$ upon entering the phase IV which can be confirmed in the panel {\bf b}. $B_\textrm{II-IV}$ (white dotted-circle), and $B_\textrm{I-IV}$ (white star) are also determined by the MCE analysis (see Figs.~\ref{fig7} and ~\ref{fig8}). The background is contour plot of $\gamma_0(x,B)$.}
\label{fig5}
\end{figure*}
%
In CeB$_6$, we observe $S_p=R\textrm{ln}2$ exactly at $T_\textrm{N}$ (Fig.~\ref{fig3}a). This indicates that, below $T_\textrm{N}$, the two lowest-lying energy levels are approximately $k_\textrm{B}T_\textrm{N}$ apart from each other while, above $T_\textrm{N}$, the level scheme transforms into two separate Kramers doublets suitable for the AFQ ordering. In Ce$_{0.82}$La$_{0.18}$B$_6$, the AFQ phase is weakened and the AFM phase is confined by a vertical phase boundary at low $B$ (Fig.~\ref{fig3}b). The isentropic line with $S_p=R\textrm{ln}2$ follows this vertical boundary validating the above mentioned fundamental change in the energy level scheme. 

The phase IV appears in Ce$_{0.77}$La$_{0.23}$B$_6$ and it is stabilized in Ce$_{0.72}$La$_{0.28}$B$_6$ with sharp $C_p$ anomalies indicating long-range ordering (Figs.~\ref{fig3}c and ~\ref{fig3}d). The I-IV boundary is almost vertical and the isentropic line with $S_p=R\textrm{ln}2$ is following this phase boundary. Whilst up to date resonant X-ray scattering (RXS) experiment~\cite{Matsumura,Matsumura1,Mannix}, NS~\cite{Kuwahara} and MF theory~\cite{Suzuki} suggest the primary order parameter is of AFO type with the AFQ and the AFM moments as field-induced secondary orders, the vertical I-IV boundary may simply indicate that the lowest-lying levels in phase IV are barely affected by the Zeeman effect. In the preliminary quantum mechanical sense, even if an isolated CEF GS is the $\Gamma_8$ quartet, exchange interactions between 15 possible order parameters could further lift the degeneracy of wave functions at each Ce-site. In reality, at $T_\textrm{I-IV}$, the level scheme seems to change into two doublets, and $S_p=R\textrm{ln}2$ along the vertical section of the I-IV boundary originates from the lower doublet. We also suggest that, when major exchange terms are taken into account, an effective $g$-factor with respect to the lowest lying doublet is very small at low $B$. 

In Ce$_{0.5}$La$_{0.5}$B$_6$, sharp $C_p$ anomalies are absent, but the above proposed properties of phase IV seem to persist below 2~\,T (Fig.~\ref{fig3}e and Fig.~\ref{fig6}). The phase IV is easily suppressed and transformed into phase I in the region where $B>1$~\,T and $T>0.25$~\,K for Ce$_{0.25}$La$_{0.75}$B$_6$ (see Fig.~\ref{fig3}f and Fig.~\ref{fig8}). 

Fig.~\ref{fig5}a shows that energy scales of AF-exchange interactions between multipolar moments are so close that most of the multipolar order parameters are competing in the limited region of the $x$-$T$ space. The temperature difference between the point where the entropy of $R\textrm{ln}2$ is released and the point where the peak in $C_p/T$ is developed (hollow square) provides an estimate for the range of critical fluctuations. As noted by the hatched area, the MF-like $C_p$ anomalies are smoothed towards the putative QCP above $x=0.75$.

In Fig.~\ref{fig5}b, $\gamma_0(x,B)$ is extracted from $C_p(T,B)/T$ by estimating constant contribution at low $T$. For $x=0$, the FL behavior is obvious and the constant value of $C_p(T,B)/T$ below 0.5~\,K (Fig.~\ref{fig1}) is taken as $\gamma_0$. As $x$ is increased, the temperature range for the FL behavior diminishes, but we can easily extract $\gamma_0(x,B)$. Even though clear FL behavior is not seen down to the lowest temperature as in phase IV ($x>0.28$), we can still extrapolate $\gamma_0$ because $C_p(T,B)/T$ shows linear $T$-dependence below 0.2~\,K. On the other hand, an applied field suppresses this $T$-linear dependence and a clear FL behavior is revealed. In this case, a part of the nuclear Schottky contribution ($\propto 1/T^3$) is included in $C_p(T,B)/T$, and we simply subtract $1/T^3$ term to get $\gamma_0$. It turns out that $\gamma_{0x}(B)$ has a local maximum at a critical point as marked by vertical dashed lines in Fig.~\ref{fig5}b. Also, the magnitude of $\gamma_0(x,B)$ drastically increases with $x$ inside phase IV. These tendencies have never been reported although resistivity measurements~\cite{Nakamura1} support strong enhancement of $m^*$ in phase IV. In contrast to the previous report~\cite{Nakamura1}, the low $T$ specific heat capacity does not support the non-Fermi-liquid (NFL) ground state. 

Symbols in Fig.~\ref{fig5}c denote critical fields determined by the MCE analysis with $T_\textrm{mix}=0.02$~\,K and the background is the contour plot of $\gamma_0(x,B)$. Representative $T_\textrm{qad}(B)$ curves with $x=0.5$ and $x=0.75$ are shown in the Figs.~\ref{fig7} and ~\ref{fig8}. Compared to the latest $x$-$B$ phase diagram which is completed by collecting largely scattered experimental results~\cite{Alistair}, the current $x$-$B$ phase diagram is more reliable in three aspects. First, critical fields are well matched with the values of the peak positions in $\gamma_{0,x}(B)$ (Fig.~\ref{fig5}b). Second, as $x$ is increased above 0.5, phase IV boundary shows negative slope towards the putative QCP. Thus, the discontinuity in the previously reported $x$-$B$ phase diagram due to the positive slope of phase IV boundary at $x>0.5$ (Ref.~\,17) is now eliminated. Thirdly, a weak phase transition between phase I and phase II is suggested (see dashed line in upper-right area of Fig.~\ref{fig5}c) which is consistent with the observation from Figs.~\ref{fig3}e and ~\ref{fig3}f that phase I will occupy the region in phase space for composition $x=0.75$ where phase II existed for $x=0.5$ (see Fig.~\ref{fig3}, Fig.~\ref{fig8} and Appendix D for more details). 

Putting aforementioned observations together, the local maximum of $\gamma_0$ at the critical point, the increase of $\gamma_0$ with $x$, and the suppression of $\gamma_0$ with $B$ strongly imply that multipolar fluctuations and $m^*$ in the heavy FL state of Ce$_{1-x}$La$_x$B$_6$ have a large positive correlation. 

As to the newly found high-entropy region below $B_\times$ in Fig.~\ref{fig5}c, we speculate that a thermally unstable (as described by the unclosed lines in Figs.~\ref{fig4}a and ~\ref{fig4}g), but weakly correlated portion of heavy electrons is segregated from the system. Such a phase segregation would allow for more electronic degrees of freedom below the black dash-dotted line in Fig.~\ref{fig5}c. Upon the mergence of the high-entropy regime with phase IV, it seems the positive correlation between fluctuating multipoles and $m^*$ is strongly enhanced.

In the near future, we hope experimental techniques such as scanning tunneling spectroscopy~\cite{Jiao}, nonresonant inelastic X-ray scattering~\cite{Sundermann} and inelastic NS~\cite{HJang} are accessible for the further characterization of both the heavy itinerant electronic state and the multipolar state in Ce$_{1-x}$La$_x$B$_6$. Also, high-quality single crystals with $0.75<x<1$ will allow us to scrutinize the existence of a putative QCP and its possible conjunction with the superconductivity which emerges as $x$ approaches to unity~\cite{Schell}. 

To conclude, a set of temperature ($T$) -- external field ($B$) phase diagrams of Ce$_{1-x}$La$_x$B$_6$ is completed by analyzing specific heat capacity, elastic neutron scattering, and the magnetocaloric effect (MCE). The $x$-$T$ phase diagram reveals competitions between multipolar energy scales at low-temperatures below 4~\,K. Local maxima of the Sommerfeld coefficient $\gamma_0$ are developed on phase boundaries while overall magnitude of $\gamma_0$ increases with $x$. The analysis of the MCE rectifies the previous $x$-$B$ phase diagram at $T$=0, and reveals a high-entropy region in lower-left corner of the modified phase diagram. The systematic change of $\gamma_0(x,B)$ on the $x$-$B$ phase diagram explicitly indicates a large positive correlation between fluctuating multipoles and the effective electron mass in the heavy Fermi-liquid state of Ce$_{1-x}$La$_x$B$_6$.

\section*{\uppercase{acknowledgments}}
We appreciate fruitful discussions with P. Thalmeier, S. Wirth and S. G{\"o}nnenwein. This work is funded by the Max Planck-POSTECH Center for Complex Phase Materials KR2011-0031558 and by the German Research Foundation (DFG) within the Research Unit 960 Quantum Phase Transitions, individual Grant No. IN 209/3-1 and the Research Training Group GRK 1621. 


\appendix
\section{Methods}

\subsection{Material preparation and characterization}
Single crystalline Ce$_{1-x}$La$_x$B$_6$ has been prepared by the floating-zone method as described elsewhere~\cite{Friemel2}. For thermodynamic measurements, we used smaller pieces of the same single crystals that were used in earlier neutron-scattering studies~\cite{Friemel1}. The structure belongs to the $Pm\overline{3}m$ space group and the lattice constant is 4.14~{\AA} for CeB$_6$. The orientation of the specimen was determined by X-ray Laue backscattering.

\subsection{Heat capacity}
The specific heat capacity is measured using a compensated heat pulse method~\cite{Whilhelm}. A reliable measurement is made in the temperature range from 0.05~\,K to 4~\,K, and the external magnetic field has been applied up to 4~\,T in the $^3$He/$^4$He dilution-fridge with 14~\,T magnet (Oxford Instrument). 

\subsection{Elastic neutron scattering}
Elastic NS measurements were performed using the cold-neutron triple-axis spectrometers IN14 at the ILL, Grenoble (Figs.~\ref{fig2}a and ~\ref{fig2}c) and the JCNS instrument PANDA~\cite{Schneidewind} at FRM-II, Garching (Figs.~\ref{fig2}b). The samples were mounted in a standard cryomagnet with the magnetic field applied along the [110] direction of the crystal. Measurements were performed using the conventional triple-axis configuration, with a cold beryllium filter on the diffracted beam to filter out higher-order contamination. The total intensity of the Bragg peak was determined by integrating the intensity either along a longitudinal scan through the $R(1/2~1/2~1/2)$ point, performed for Fig~\ref{fig2}a, or through the integration of a rocking scan through the same point in Figs.~\ref{fig2}b and ~\ref{fig2}c.

\subsection{Magnetocaloric effect}
Quasi-adiabatic condition were created by supporting the sample holder with thin nylon wires. The relaxation of the sample temperature to the temperature of the mixing chamber of the dilution-fridge is well described by a heat transfer equation and the relaxation time is about 4 hours from 1~\,K. Therefore, a sweep-rate has been chosen so that the total period of a field-sweep is shorter than 4 hours. Under the above conditions, the sample temperature is monitored as a function of the external magnetic field. 


\begin{figure}[h]
\includegraphics[width=\columnwidth]{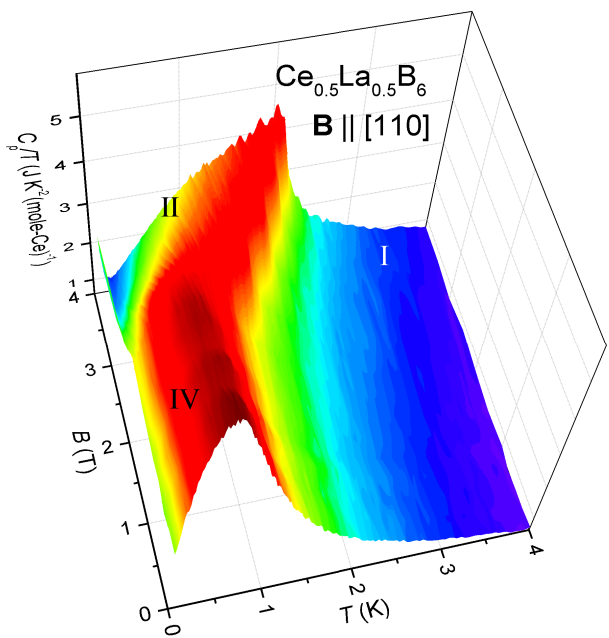}
\caption{ {\bf(color) Surface plot of the specific heat capacity divided by temperature obtained from Ce$_{0.5}$La$_{0.5}$B$_6$.} External magnetic field is applied along [110] direction and regions of phases I, II, and IV are briefly noted.}
\label{fig6}
\end{figure} 


\section{Representative Surface plot of the specific heat capacity}
Specific heat capacity under constant pressure and field, $C_{p,B}$, of Ce$_{0.5}$La$_{0.5}$B$_6$ is measured as a function of temperature. The external magnetic fields has been varied from 0~\,T to 4~\,T with 0.25~\,T of interval. Then, the obtained $C_{p,B}(T)$ curves compose the framework for the surface plot. Field-insensitive peaks around 1~\,K and below 2.5~\,T are clearly shown and we suspect these peaks indicate rather broadened phase transition between phase I and phase IV. As the field is further increased, phase IV is suppressed and phase II emerges. This can be recognized by the field-induced (1/2 1/2 1/2) Bragg intensity (not shown). Phase transitions are not very clear in 2D plots but these are more distinguishable in the 3D plot (Fig.~\ref{fig6}).


\begin{figure*}[h]
\includegraphics[width=\textwidth]{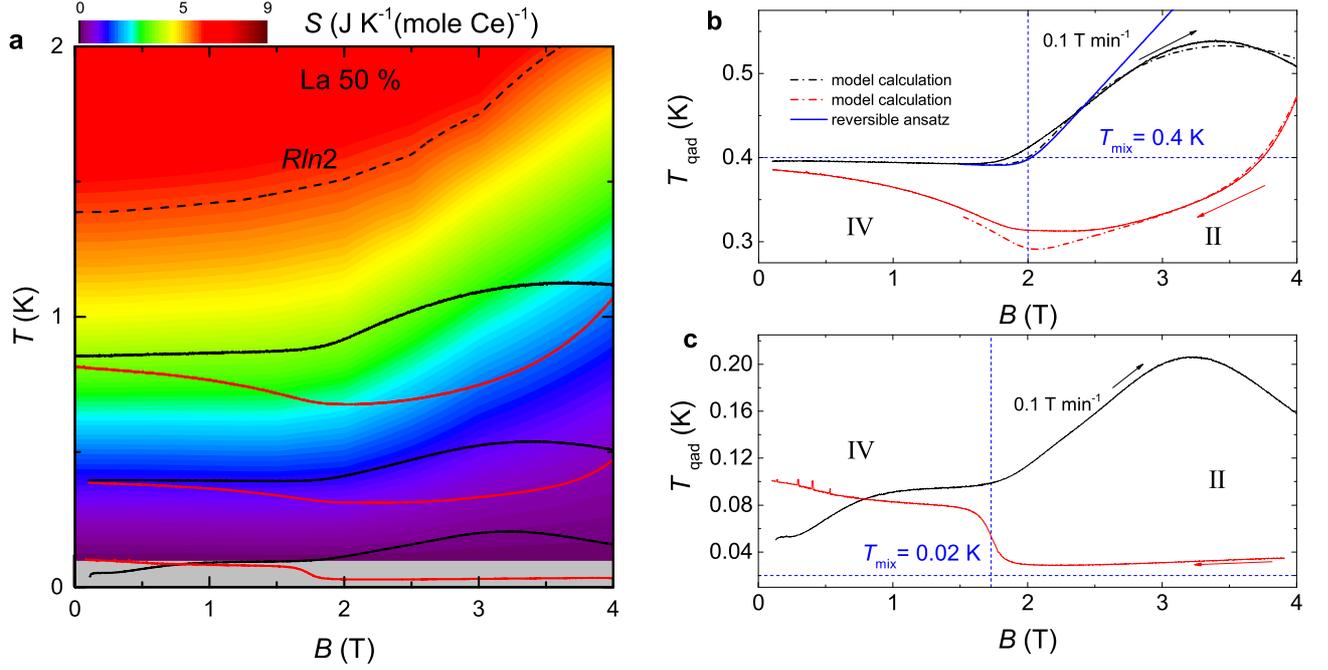}
\caption{ {\bf Quasi-adiabatic MCE in Ce$_{0.5}$La$_{0.5}$B$_6$ superimposed on the contour plot of its entropy.} ({\bf a}) Representative $T_\textrm{qad}(B)$ curves with $T_\textrm{mix}=0.02$~\,K, 0.4~\,K, and 0.8~\,K are shown. For a reference, $S_p=R\textrm{ln}2$ line is inserted with the black dashed-line and the background is the contour plot of the entropy. Black and red solid-lines are results with increasing and decreasing field, respectively. ({\bf b}) $T_\textrm{qad}(B)$ curves with $T_\textrm{mix}=0.4$~\,K are displayed. The model calculations without field-induced heating are exhibited with dash-dotted lines. The blue vertical dashed-line is placed to demarcate the phase II-IV transition. ({\bf c}) $T_\textrm{qad}(B)$ curves with $T_\textrm{mix}=0.02$~\,K are displayed. The blue vertical dashed-line is placed where $\abs{dT_\textrm{qad}(B)/dB}$ is the largest in down-sweep. The sweep-rate is 0.1 T min$^{-1}$ in all panels.}
\label{fig7}
\end{figure*} 


\begin{figure*}[h]
\includegraphics[width=\textwidth]{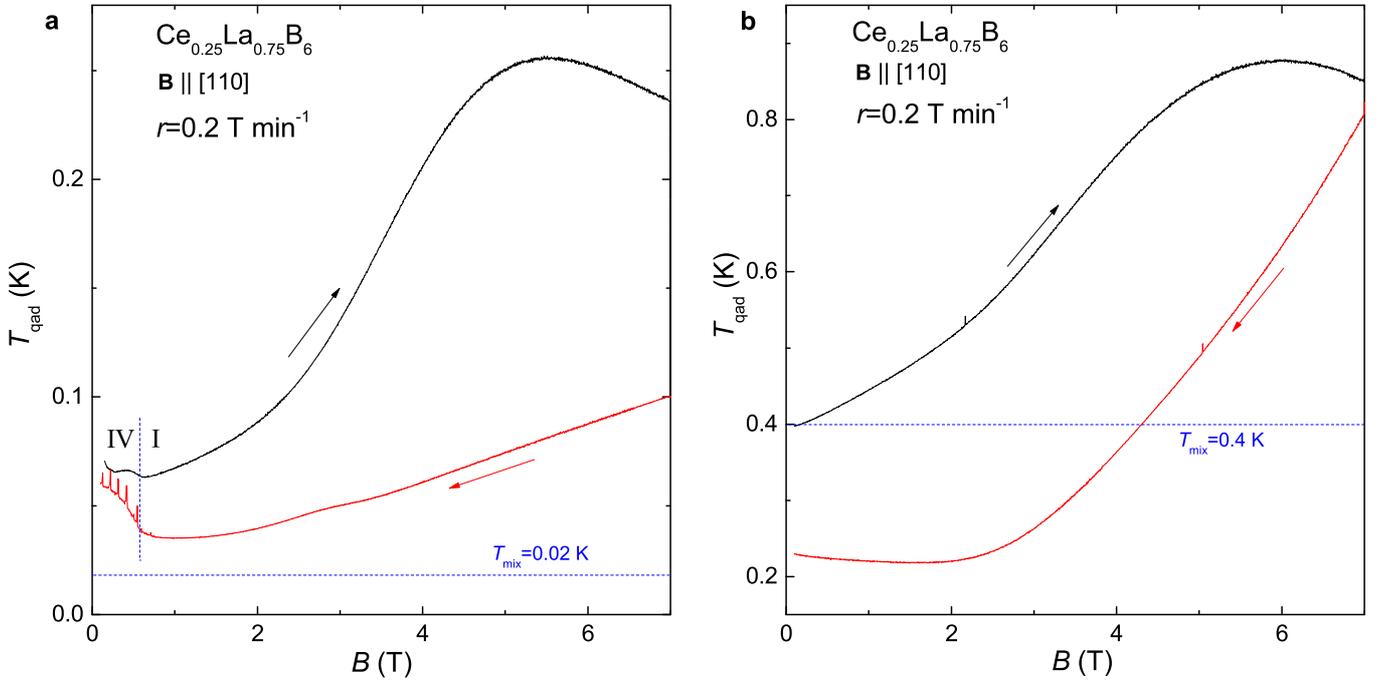}
\caption{ {\bf Quasi-adiabatic MCE observed in Ce$_{0.25}$La$_{0.75}$B$_6$.} ({\bf a}) Black and red solid-lines are $T_\textrm{qad}(B)$ curves with increasing and decreasing field, respectively. The temperature of the mixing chamber $T_\textrm{mix}$ is 0.02~\,K. The blue vertical dashed line marks the phase transition between the phase I and the phase IV. ({\bf b}) Curves for the quasi-adiabatic temperature $T_\textrm{qad}(B)$ are shown at $T_\textrm{mix}=0.4$~\,K. The sweep-rate is 0.2~\,T min$^{-1}$ in both panels.}
\label{fig8}
\end{figure*} 


\section{Analysis of the quasi-adiabatic MCE}
Here, we derive the differential equation for the theoretical quasi-adiabatic temperature, $T_\textrm{qad}^0(B)$. First, we consider a term for the genuine (reversible) MCE, and then add a term which reflects an excessive field-induced heating. Given that the specimen with a virtual temperature $T_\textrm{qad}^0$ is linked to the mixing chamber at the temperature of $T_\textrm{mix}$, there should also be a term representing the natural relaxation of $T_\textrm{qad}^0$ to $T_\textrm{mix}$. Taking pieces together, we can write the general differential equation with respect to $dT_\textrm{qad}^0/dB$ as 

\begin{equation}
\begin{aligned}
\frac{dT_\textrm{qad}^0}{dB}=& \frac{dT_\textrm{rev}}{dB} \pm \frac{dT_\textrm{irr}^\pm}{dB} \\ 
& \mp \frac{1}{\abs{r}C_p(T,B)}\int_{T_\textrm{mix}}^{T_\textrm{qad}^0} dT K(T,B),
\end{aligned}
\end{equation}

where the double signs are in the same order and the equation with the upper (lower) signs is for the increasing (decreasing) field. $dT_\textrm{rev}$ is an infinitesimal temperature change following an isentropic line. $dT_\textrm{irr}^\pm$ is an infinitesimal temperature change due to a field-induced heating other than an eddy current heating. It is quantitatively discussed in the last paragraph of the current section that the eddy current heating is negligible. The thermal conduction coefficient $K(T,B)$ between the sample and the mixing chamber is extracted by analyzing the relaxation behavior of $T_{\textrm{qad},B}(t)$ to the $T_\textrm{mix}$, and we found $K(T,B)=\alpha T^\delta$ with negligible $B$-dependence below 5~\,T. Here, $\alpha\simeq 100$~\,nW K$^{-1}$ and $\delta$=1.2. $r$ in the last term is the field sweep-rate.

By experience, we already know that many of analytic expressions for thermodynamic functions are written by various combinations of hyperbolic functions. Hence, $T_\textrm{rev} \propto \{\mathrm{tanh}(B-B_\textrm{c})+1\}$ provides a good starting point to construct a reversible ansatz for the MCE since this function well depicts stepwise temperature change upon the phase transition. Here, $\abs{dT_\textrm{qad}^0/dB}$ is the largest at an arbitrary critical field $B_\textrm{c}$. We further elaborate this idea to imitate various shapes of isentropic lines calculated in the main text with the following expression,

\begin{equation}
\begin{aligned}
T_\textrm{rev}(B_\textrm{f})=& C_1^\textrm{rev}\{\mathrm{tanh}(w_1^\textrm{rev}(B_\textrm{f}-B_\textrm{c}))+1\}\\ 
& +\frac{C_2^\textrm{rev}}{w_2^\textrm{rev}}\mathrm{ln}\abs{\frac{\mathrm{cosh}\{w_2^\textrm{rev} (B_\textrm{f}-B_\textrm{c})\}}{\mathrm{cosh}\{w_2^\textrm{rev} (B_\textrm{i}-B_\textrm{c})\}}}\\
& +w_3^\textrm{rev} C_2^\textrm{rev} (B_\textrm{f}-B_\textrm{i}),
\end{aligned}
\end{equation}
 
where $C_1^\textrm{rev}$ determines the size of the stepwise MCE in a narrow range of field close to $B_\textrm{c}$, and $w_1^\textrm{rev}$ determines sharpness of the transition. Shapes of isentropic lines in a wider range of field are determined by adjusting $C_2^\textrm{rev}$, $w_2^\textrm{rev}$, and $w_3^\textrm{rev}$. All the blue curves for the reversible ansatz in Fig.~\ref{fig4} of the main text are generated by eq. (2). $B_\textrm{i}$ and $B_\textrm{f}$ are the initial and the final fields for the eq. (1), respectively.  

The irreversible temperature change is approximated by the following function,  

\begin{equation}
\begin{aligned}
T_\textrm{irr}^\pm(B)= \sum_{i=1}^2 D_i^\pm \{\mathrm{tanh}(w_i^\pm(B-B_i^\pm)+1\}.
\end{aligned}
\end{equation} 

The envelopes of red-filled and black-hatched areas in Fig.~\ref{fig4} are described by the relation $dQ/dB \simeq C_p(T_\textrm{qad},B)dT_\textrm{irr}^\pm(B)/dB$. Finally, eqs. (2) and (3) are implanted in eq. (1) and numerically solved for $T_\textrm{qad}^0(B)$. The parameters in eqs. (2) and (3) are iteratively adjusted until the $T_\textrm{qad}^0(B)$ is optimized to the $T_\textrm{qad}(B)$. It should be emphasized that more than two hyperbolic terms in eq. (3) do not improve the quality of the numerical calculation. Also, note that we are dealing with only the second order phase transition because the ansatz for the genuine MCE is continuous and reversible. 

Here, we do not enumerate values of parameters but summarize specific conditions which are unique to different of phase boundaries. In case $B_1^{+}$=$B_2^{+}$=$B_1^{-}$=$B_2^{-}$=$B_\textrm{c}$ and $D_i^{+}$=$D_i^{-}$, the heating curves are reversible and II-III$^\prime$ transitions are well described by this condition. When $B_1^+$=$B_2^+ \neq B_1^-$=$B_2^-$ and $D_i^+$=$D_i^-$ AFM structural phase transitions around III-III$^\prime$ boundaries are well reproduced. In case $B_1^+$=$B_2^+\neq B_1^-$=$B_2^-$ and $D_i^+\neq D_i^-$, the $dQ/dB$ curves can have different shapes depending on the sweep direction. Therefore we can describe highly hysteretic features in $T_\textrm{qad}(B)$.

It is suggested that an inelastic scattering of quasiparticles triggered by strongly fluctuating local moments could be the major cause of the irreversible $T_\textrm{qad}(B)$ in the vicinity of a phase transition. We designate this sort of field-induced heating as the critical heating. The magnitude of the heating due to the domain motion, which we call the domain heating, is usually much smaller than the magnitude of the critical heating, and the domain heating is not concentrated in a narrow range around a phase boundary. This fact is well represented in Fig.~\ref{fig4}b. In phase III$^\prime$, highly populated magnetic domains are expected from NS, and we do observe substantial domain heating: $T_\textrm{qad}(B)$ is monotonically increasing over wide the range of field regardless of the sweep direction. 

Power dissipation per unit volume by the eddy current is given by $P/V = E_{\phi}(R)^{2}/2\rho = R^{2}(\partial B_{z}/\partial t)^{2}/8\rho$ assuming a cylindrical geometry of radius $R$, height $L$. The field is along the $z$-direction. Taking the average dimension of the specimen $R$=$L$=1~\,mm and the resistivity $\rho \simeq$10~\,$\mu\Omega$cm, 0.26~\,nJ of eddy current heating $Q_\textrm{eddy}$ is generated for 40 min with $r=\partial B_{z}/\partial t=0.1$~\,T min$^{-1}$. At 0.1~\,K, the heat capacity, $C$, of 5~\,mg of CeB$_6$ is 0.37~\,$\mu$JK$^{-1}$. From these values, the temperature increase $\Delta T=Q_\textrm{eddy}/C$ is estimated as 0.7~\,mK for 40 min in a perfectly adiabatic condition. It is definitely negligible compared to notable changes in $T_\textrm{qad}(B)$.

\section{Empirical Determination of critical points from the MCE}

At $x=0.5$, the quasi-adiabatic MCE shows a broad transition feature between phase II and phase IV except at very low-$T$ below 0.05~\,K (Fig.~\ref{fig7}a). For $T_\textrm{mix}=$0.4~\,K, eq. (3) is solved without an excessive heating term and solutions are superimposed on $T_\textrm{qad}(B)$ curves with dash-dotted lines as shown in the Fig.~\ref{fig7}b. The steepest slope of the reversible ansatz is assumed to appear at the critical field as noted by the vertical blue dashed-line. On the other hand, the critical heating strongly affects the $T_\textrm{qad}(B)$ at very low $T$: See the down-sweep curve in Fig.~\ref{fig7}c. In this case, the numerical calculation is not possible because we cannot estimate correct values for $C_p(T,B)$ below 0.05~\,K, and the critical point is located where $\abs{dT_\textrm{qad}(B)/dB}$ is the largest. It is conjectured that the MC sweep correctly captures intensive fluctuations of order parameters in the vicinity of critical points noted in Fig.~\ref{fig5}c.

Fig.~\ref{fig8} exhibits $T_\textrm{qad}(B)$ curves in Ce$_{0.25}$La$_{0.75}$B$_6$. When the mixing chamber is anchored to 0.02~\,K, we observe the critical heating below 0.8~\,T and the star-shaped symbol in Fig.~\ref{fig5}c is referenced to this value. We deduce that the quantum fluctuation regarding the I-IV phase transition at 0.8~\,T is detected by the MC sweep. This statement is strongly supported by the observation of the local maximum in $\gamma_0$ at the same field (Fig.~\ref{fig5}b). In addition, it must be noted that there exists a subtle wiggling in $T_\textrm{qad}(B)$ around 3.5~\,T below 0.05~\,K. This is suspected as reminiscent of the weak I-II transition. As the system is away from $T=0$, the critical heating is no longer observed and only the paramagnetic response is remained (Fig.~\ref{fig8}b). The spikes in the red curve is due to the flux jump from the superconducting magnet. 
  



\begin{thebibliography}{10}

\bibitem{Zirngiebl}
E. Zirngiebl, B. Hillebrands, S. Blumenr{\"o}der, G. G{\"u}ntherodt, M. Loewenhaupt, J. M. Carpenter, K. Winzer, and Z. Fisk, Phys. Rev. B {\bf 30}, 4052 (1984).

\bibitem{Ohkawa1}
J. F. Ohkawa, J. Phys. Soc. Jpn. {\bf 52}, 3897 (1983).

\bibitem{Ohkawa2}
J. F. Ohkawa, J. Phys. Soc. Jpn. {\bf 54}, 3909 (1985).

\bibitem{Shiina}
R. Shiina, H. Shiba, and P. Thalmeier, J. Phys. Soc. Jpn. {\bf 66}, 1741 (1997).

\bibitem{Effantin}
J. M. Effantin, P. Burlet, J. Rossat-Mignod, S. Kunii, and T. Kasuya, Proc. Int. Conf. Valence Instabilities (Zurich, 1982) ed. P. Wachter and H. Boppart (Amsterdam: North-Holland).

\bibitem{Hiroi}
M. Hiroi, S. Kobayashi, M. Sera, N. Kobayashi, and S. Kunii, J. Phys. Soc. Jpn. {\bf66}, 1762 (1997).

\bibitem{Zaharko}
O. Zaharko, P. Fischer, A. Schenk, S. Kunii, P.-J. Brown, F. Tasset, and T. Hansen, Phys. Rev. B {\bf68}, 214401 (2003).

\bibitem{Sera}
M. Sera, H. Ichikawa, T. Yokoo, J. Akimitsu, M. Nishi, K. Kakurai, and S. Kunii, Phys. Rev. Lett. {\bf 86}, 1578 (2001).

\bibitem{Sera1}
M. Sera, and S. Kobayashi, J. Phys. Soc. Jpn. {\bf68}, 1664 (1999).

\bibitem{Nakao}
H. Nakao, K. Magishi, Y. Wakabayashi, Y. Murakami, K. Koyama, K. Hirota, Y. Endoh S. Kunii, J. Phys. Soc. Jpn. {\bf70}, 1857 (2001).

\bibitem{Takigawa}
M. Takigawa, H. Yasuoka, T. Tanaka, and Y. Ishizawa, J. Phys. Soc. Jpn. {\bf 52}, 728 (1983).

\bibitem{Sakai}
O. Sakai, R. Shiina, H. Shiba, and P. Thalmeier, J. Phys. Soc. Jpn. {\bf 66}, 3005 (1997).

\bibitem{Matsumura}
T. Matsumura, T. Yonemura, K. Kunimori, M. Sera, F. Iga, T. Nagao, and J. I. Igarashi, Phys. Rev. B {\bf 85}, 174417 (2012). 

\bibitem{Matsumura1}
T. Matsumura, S. Michimura, T. Inami, T. Otsubo, H. Tanida, F. Iga, and M. Sera, Phys. Rev. B {\bf 89}, 014422 (2014).

\bibitem{Mannix}
D. Mannix, Y. Tanaka, D. Carbone, N. Bernhoeft, and S. Kunii, Phys. Rev. Lett. {\bf 95}, 117206 (2005).

\bibitem{Kuwahara}
K. Kuwahara, K. Iwasa, M. Kohgi, N. Aso, M. Sera, and F. Iga, J. Phys. Soc. Jpn. {\bf 76}, 093402 (2007).

\bibitem{Alistair}
A. S. Cameron, G. Friemel, and D. S. Inosov, Rep. Prog. Phys. {\bf79}, 066502 (2016).

\bibitem{Friemel}
G. Friemel, \textit{`Itinerant spin dynamics in iron-based superconductors and cerium-based heavy-fermion antiferromagnets'}, Ph. D. Thesis, Fakult{\"a}t Mathematik und Physik, Universit{\"a}t Stuttgart (2014).

\bibitem{Nakamura}
S. Nakamura, T. Goto, O. Suzuki, S. Kunii, and S. Sakatsume, Phys. Rev. B {\bf 61}, 15203 (2000).

\bibitem{Tayama}
T. Tayama, T. Sakakibara, K. Tenya, H. Amitsuka and S. Kunii, J. Phys. Soc. Jpn. {\bf 66}, 2268 (1997).

\bibitem{Pavlo}
P. Y. Portnichenko, S. Paschen, A. Prokofiev, M. Vojta, A. S. Cameron, J.-M. Mignot, A. Ivanov, and D. S. Inosov, Phys. Rev. B {\bf 94}, 245132 (2016).

\bibitem{Schotte}
K. D. Schotte and U. Schotte, Phys. Lett. {\bf 55A}, 38 (1975).

\bibitem{Desgranges}
H. U. Destgranges and K. D. Schotte, Phys. Lett. {\bf 91A}, 240 (1982). 

\bibitem{Nakamura2}
S. Nakamura, T. Goto, and S. Kunii, J. Phys. Soc. Jpn. {\bf 64}, 3941 (1995).

\bibitem{Akatsu}
M. Akatsu, T. Goto, Y. Nemeto, O. Susuki, S. Nakamura, and S. Kunii, J. Phys. Soc. Jpn. {\bf 72}, 205 (2003).

\bibitem{Suzuki}
O. Suzuki, T. Goto, S. Nakamura, T. Matsumura, and S. Kunii, J. Phys. Soc. Jpn, {\bf 67}, 4243 (2007). 

\bibitem{Nakamura1}
S. Nakamura, M. Endo, H. Yamamoto, T. Isshiki, N. Kimura, H. Aoki, T. Nojima, S. Otani, and S. Kunii, Phys. Rev. Lett. {\bf 97}, 237204 (2007).

\bibitem{Friemel1}
G. Friemel, H. Jang, A. Schneidewind, A. Ivanov, A. V. Dukhnenko, N. Y. Shitsevalova, V. B. Filipov, B. Keimer, and D. S. Inosov, Phys. Rev. B. {\bf 92}, 014410 (2015). 

\bibitem{Jiao}
L. Jiao, S. R{\"o}{\ss}ler, D. J. Kim, L. H. Tjeng, Z. Fisk, F. Steglich, and S. Wirth, Nat. Commun. {\bf 7}, 1 (2016). 

\bibitem{Sundermann}
M. Sundermann, M. W. Haverkort, S. Agrestini, A. Al-Zein, M. Moretti Sala, Y. Huang, M. Golden, A. de Visser, P. Thalmeier, L. H. Tjeng, A. Severing, Proc. Natl. Acad. Sci. U. S. A. {\bf 113}, 13989 (2016). 


\bibitem{HJang}
H. Jang, G. Friemel, J. Ollivier, A. V. Dukhnenko, N. Y. Shitsevalova, V. B. Filipov, B. Keimer, and D. S. Inosov, Nat. Mater. {\bf 13}, 682 (2014).

\bibitem{Schell}
G. Schell, H. Winter, H. Rietschel, and F. Gompf, Phys. Rev. B {\bf 25}, 1589 (1982).

\bibitem{Friemel2}
G. Friemel, Y. Li, A. V. Dukhnenko, N. Y. Shitsevalova, N. E. Sluchanko, A. Ivanov, V. B. Filipov, B. Keimer, and D. S. Inosov, Nat. Commun. 3, 830 (2012).

\bibitem{Whilhelm}
H. Wilhelm, T. L{\"u}hmann, T. Rus, and F. Steglich, Rev. Sci. Instrum. {\bf 75}, 2700, (2004).

\bibitem{Schneidewind}
A. Schneidewind and P. \v{C}erm\'ak, J. Large-Scale Res. Facilities {\bf 1}, A12 (2015). doi:10.17815/jlsrf-1-35



\end{thebibliography}
\end{document}